\documentclass[%
 reprint,
superscriptaddress,
 amsmath,amssymb,
 aps,
]{revtex4-2}

\usepackage{graphicx}
\usepackage{dcolumn}
\usepackage{bm}
\usepackage{xcolor}
\usepackage[colorlinks=true, linkcolor=blue, citecolor=red, urlcolor=magenta]{hyperref}
\usepackage{microtype}
\usepackage{orcidlink}

\begin{document}


\title{Saha's ionization in the Schwarzschild spacetime}

\author{Lázaro L. Sales\,\orcidlink{0000-0002-5352-6642}}
 \email{lazaro0254@gmail.com}
 \affiliation{Departamento de F\'{\i}sica, Universidade Federal de Campina Grande, Caixa Postal 10071, 58429-900, Campina Grande, Para\'{\i}ba, Brazil}
 
\author{Jonatas A. Silva\,\orcidlink{0000-0003-2246-1660}}
 \email{arizilanio@hotmail.com}
 \affiliation{Departamento de F\'isica, Universidade do Estado do Rio Grande do Norte, 59610-210, Mossor\'o-RN, Brazil}

 \author{Amilcar R. Queiroz\,\orcidlink{0000-0002-4785-5589}}
 \email{amilcarq@df.ufcg.edu.br}
 \affiliation{Departamento de F\'{\i}sica, Universidade Federal de Campina Grande, Caixa Postal 10071, 58429-900, Campina Grande, Para\'{\i}ba, Brazil}

\date{\today}

\begin{abstract}
We investigate the Saha ionization equilibrium in the exterior Schwarzschild spacetime. Starting from the mass-shell condition for a massive particle, we derive the conserved energy associated with the timelike Killing symmetry and obtain its nonrelativistic limit in terms of the Schwarzschild lapse function. This energy is then incorporated into the Maxwell--Boltzmann distribution to construct the ionization balance as seen by an asymptotic observer. We show that the gravitational field modifies the Saha relation through the redshift of both the particle energies and the thermodynamic variables. By imposing the Tolman--Ehrenfest and Klein equilibrium conditions, the formulation expressed entirely in terms of locally measured quantities reduces to the standard flat-spacetime Saha equation. We further analyze the gravitational redshift of the ionization energy, the relative departure from the flat-spacetime ionization ratio, and the corresponding shift of the ionization fraction as a function of the asymptotic temperature. We also present the relativistic classical counterpart based on the Maxwell--J\"uttner distribution and show explicitly that its leading correction is negligible in the temperature range where neutral hydrogen is appreciably abundant. These results provide a consistent description of ionization equilibrium in a static gravitational field and clarify the distinction between gravitational redshift effects and intrinsic modifications of local atomic physics.
\end{abstract}

\maketitle


\section{Introduction}
\label{sec:introduction}

The ionization state of a dilute gas is determined by the interplay between its microscopic atomic structure and the thermodynamic conditions of the surrounding medium. Under the assumptions of thermal and chemical equilibrium, this relation is described by the Saha equation, originally developed in the context of stellar atmospheres \cite{Saha1920} and subsequently applied to a broad range of problems in astrophysics, plasma physics, and cosmology \cite{gangopadhyay2021hundred,dodelson2024modern}. In particular, the Saha formalism provides the equilibrium limit of primordial hydrogen recombination, although an accurate description of the full recombination history requires kinetic treatments that account for departures from equilibrium, radiative transfer, and transitions through excited atomic states \cite{peebles1968recombination,seager1999new,chluba2006free,chluba2011towards,dubrovich2005recombination}.

The equilibrium structure underlying the Saha equation also provides a useful framework for investigating how ionization processes respond to modifications of particle statistics, chemical potentials, and single-particle energies. Previous studies have explored non-Gaussian corrections to primordial ionization \cite{sales2022non}, the evolution of the chemical potentials of the reacting species \cite{sales2023constraint}, phenomenological descriptions of departures from equilibrium \cite{sales2025possible}, and the formulation of the Saha equation in uniformly accelerated frames \cite{sales2024non}. These results indicate that the ionization balance is particularly sensitive to the energy entering the equilibrium distribution and to the thermodynamic variables used by the observer.

In a static gravitational field, however, energy and temperature must be defined with particular care. The energy measured by a local observer generally differs from the conserved Killing energy associated with the timelike symmetry of the spacetime. Likewise, global thermal equilibrium does not imply a spatially uniform local temperature: in a static geometry, the local temperature is related to the gravitational redshift through the Tolman--Ehrenfest relation \cite{tolman1930temperature,tolman1935thermal,Santiago_2019}. The corresponding redshift relation for the chemical potential is commonly expressed through the Klein condition \cite{klein1949thermodynamical,lima2019thermodynamic}. These relations are essential for constructing a thermodynamically consistent ionization equation in curved spacetime.

In this work, we formulate the Saha ionization equation in the exterior Schwarzschild geometry. Starting from the mass-shell condition, we derive the conserved relativistic energy of a massive particle and obtain its nonrelativistic limit in terms of the Schwarzschild lapse function. We then use this energy in the Maxwell--Boltzmann distribution to derive the ionization balance as expressed through thermodynamic quantities referred to an observer at spatial infinity. Finally, by imposing the Tolman--Ehrenfest and Klein equilibrium relations, we show that the equation written entirely in terms of locally measured quantities recovers the standard flat-spacetime Saha form. Thus, the gravitational field affects the relation between local and asymptotic descriptions through gravitational redshift, while the intrinsic ionization energy remains unchanged in a sufficiently small local inertial frame, consistent with the equivalence principle.

The remainder of this paper is organized as follows. In Sec.~\ref{sec:saha}, we briefly review the standard Saha ionization equation and summarize the assumptions underlying its equilibrium formulation. In Sec.~\ref{sec:schwarzschild_energy}, we derive the conserved energy of a massive particle in the Schwarzschild spacetime and discuss its relation to the locally measured energy. In Sec.~\ref{sec:results}, we construct the Saha equation in the Schwarzschild geometry using the corresponding nonrelativistic energy and analyze its interpretation in terms of local and asymptotic thermodynamic quantities. We then derive its relativistic classical counterpart, recover the nonrelativistic result through the large-argument expansion of the modified Bessel function, and estimate the leading kinematic correction. Finally, our conclusions are presented in Sec.~\ref{sec:conclusion}.

\section{Saha's ionization}
\label{sec:saha}

The Saha ionization equation provides the equilibrium relation between the abundances of neutral atoms, ions, and free electrons in a dilute plasma. Originally introduced to describe ionization processes in stellar atmospheres, it has become a fundamental tool in astrophysics, plasma physics, and cosmology. In the context of the early Universe, it offers a simple description of the onset of primordial hydrogen recombination, relating the ionization fraction to the temperature of the photon bath under the assumptions of local thermodynamic and chemical equilibrium \cite{Saha1920,dodelson2024modern,peebles2020principles}.

For the reaction
\begin{equation}
e^- + p
\rightleftharpoons
H_n+\gamma,
\label{eq:reaction}
\end{equation}
chemical equilibrium requires
\begin{equation}
\mu_e+\mu_p=\mu_{H_n},
\label{eq:chemical_equilibrium}
\end{equation}
where the photon chemical potential vanishes. Assuming that the particles obey Maxwell--Boltzmann statistics, the number density of each nonrelativistic species is given by \cite{dodelson2024modern}
\begin{equation}
n_i
=
\frac{g_i}{(2\pi\hbar)^3}
\int d^3p\,
\exp\!\left[
-\frac{E_i-\mu_i}{k_{\rm B}T}
\right],
\label{eq:number_density_MB}
\end{equation}
with \(g_i\), \(E_i\), and \(\mu_i\) denoting the degeneracy factor, single-particle energy, and chemical potential, respectively.

Evaluating the momentum integral in the nonrelativistic limit,
\begin{equation}
E_i
=
m_ic^2+\frac{p^2}{2m_i},
\label{eq:nonrel_energy}
\end{equation}
one finds
\begin{equation}
n_i
=
g_i
\left(
\frac{m_i k_{\rm B}T}
{2\pi\hbar^2}
\right)^{3/2}
\exp\!\left[
\frac{\mu_i-m_ic^2}
{k_{\rm B}T}
\right].
\label{eq:number_density}
\end{equation}

Combining Eq.~(\ref{eq:number_density}) with the equilibrium condition (\ref{eq:chemical_equilibrium}), the Saha equation assumes the familiar form
\begin{equation}
\frac{n_pn_e}{n_{H_n}}
=
G_{(e,p,n)}
\left(
\frac{m_{\rm red}k_{\rm B}T}
{2\pi\hbar^2}
\right)^{3/2}
\exp\!\left(
-\frac{\varepsilon_n}
{k_{\rm B}T}
\right),
\label{eq:saha_standard}
\end{equation}
where
\begin{equation}
m_{\rm red}
=
\frac{m_em_p}{m_e+m_p}
\end{equation}
is the reduced mass,
\begin{equation}
\varepsilon_n
=
(m_p+m_e-m_{H_n})c^2
\end{equation}
is the $n$-th excited state of the hydrogen atom (or ionization potential), and
\begin{equation}
G_{(e,p,n)}
=
\frac{g_eg_p}{g_{H_n}}
\end{equation}
is the ratio of statistical weights.

Equation (\ref{eq:saha_standard}) is only valid while the plasma remains in thermodynamic equilibrium. During cosmological hydrogen recombination, however, equilibrium gradually breaks down because electrons are predominantly captured into excited states, followed by radiative cascades toward the ground state \cite{peebles2020principles}. These processes are described more accurately by kinetic approaches based on the Boltzmann equation, such as the Peebles model \cite{peebles1968recombination} and modern recombination codes \cite{lee2020hyrec,chluba2011towards}. Consequently, the Saha equation provides only an approximate description of the early stages of recombination, although it remains an essential theoretical reference for equilibrium analyses \cite{gangopadhyay2021hundred}.

In recent years, several extensions of the standard Saha formalism have been investigated (see, e.g., \cite{sales2022non,sales2024non,sales2025possible}). These developments demonstrate that the Saha equation can be naturally generalized whenever the underlying particle energy or the equilibrium distribution is modified while preserving the thermodynamic structure of the formalism. The Schwarzschild extension developed in the following section follows the same philosophy by replacing the flat-spacetime particle energy with its corresponding expression in a static curved spacetime.

\section{Energy in the Schwarzschild spacetime}
\label{sec:schwarzschild_energy}

In this section, we review the main properties of the Schwarzschild geometry that will be required throughout this work. In particular, we derive the conserved energy associated with the timelike symmetry of the spacetime and distinguish it from the energy measured by a static local observer. This distinction provides the physical basis for the formulation of the Saha equation developed in Sec. \ref{sec:saha_schwarzschild}.

The Schwarzschild solution describes the exterior gravitational field generated by a static and spherically symmetric body of mass \(M\). In spherical coordinates \((t,r,\theta,\varphi)\), the line element is
\begin{equation}
ds^{2}
=
-f(r)c^{2}dt^{2}
+
\frac{dr^{2}}{f(r)}
+
r^{2}d\theta^{2}
+
r^{2}\sin^{2}\theta\,d\varphi^{2},
\label{eq:schwarzschild_metric_saha}
\end{equation}
where
\begin{equation}
f(r)
=
1-\frac{2GM}{rc^{2}}.
\label{eq:schwarzschild_function}
\end{equation}
It is convenient to introduce the Schwarzschild lapse function,
\begin{equation}
N(r)
\equiv
\sqrt{f(r)}
=
\sqrt{1-\frac{2GM}{rc^{2}}}.
\label{eq:schwarzschild_lapse}
\end{equation}

Since the metric coefficients are independent of the time coordinate \(t\), the Schwarzschild spacetime is stationary. It therefore admits the timelike Killing vector
\begin{equation}
K^{\mu}
=
(1,0,0,0),
\label{eq:killing}
\end{equation}
which generates time translations and gives rise to a conserved quantity along geodesic motion.

For a particle with four-momentum \(p^{\mu}\), the conserved quantity associated with the Killing vector is
\begin{equation}
\mathcal{E}
=
-K_{\mu}p^{\mu},
\label{eq:killing_energy_general}
\end{equation}
where
\begin{equation}
K_{\mu}
=
g_{\mu\nu}K^{\nu}.
\label{eq:killing_cov}
\end{equation}
Using the Schwarzschild metric, one finds
\begin{equation}
K_{\mu}
=
\left(
-f(r),
0,
0,
0
\right),
\label{eq:killing_cov_schwarzschild}
\end{equation}
when the temporal coordinate is written as \(x^{0}=ct\). Consequently, the conserved physical energy associated with the Schwarzschild time coordinate can be written as
\begin{equation}
E_{\infty}
\equiv
-cp_{0}.
\label{eq:energy_infinity_definition}
\end{equation}
This quantity is commonly referred to as the Killing energy. It is conserved along the trajectory of a freely falling particle and coincides with the energy measured by an asymptotic observer in the limit \(r\rightarrow\infty\).

Following the procedure adopted for the Rindler spacetime in Ref.~\cite{sales2024non}, it is important to distinguish the conserved energy \(E_{\infty}\) from the energy \(E_{\rm loc}\) measured by a static observer located at a finite radial coordinate \(r\). The former is defined globally through the timelike Killing symmetry, whereas the latter corresponds to the energy measured in the observer's local orthonormal frame.

The four-momentum of a particle of rest mass \(m\) satisfies the mass-shell condition
\begin{equation}
g^{\mu\nu}p_{\mu}p_{\nu}
=
-m^{2}c^{2}.
\label{eq:mass_shell_schwarzschild}
\end{equation}
From Eq.~\eqref{eq:schwarzschild_metric_saha}, this relation becomes
\begin{equation}
-\frac{p_{0}^{2}}{f(r)}
+
f(r)p_{r}^{2}
+
\frac{p_{\theta}^{2}}{r^{2}}
+
\frac{p_{\varphi}^{2}}{r^{2}\sin^{2}\theta}
=
-m^{2}c^{2}.
\label{eq:mass_shell_explicit}
\end{equation}

The squared physical spatial momentum measured by a static local observer is defined as
\begin{equation}
p_{\rm loc}^{2}
\equiv
f(r)p_{r}^{2}
+
\frac{p_{\theta}^{2}}{r^{2}}
+
\frac{p_{\varphi}^{2}}{r^{2}\sin^{2}\theta}.
\label{eq:local_momentum}
\end{equation}
Using Eq.~\eqref{eq:energy_infinity_definition}, the mass-shell relation yields
\begin{equation}
E_{\infty}
=
N(r)
\sqrt{
m^{2}c^{4}
+
p_{\rm loc}^{2}c^{2}
}.
\label{eq:energy_infinity_exact}
\end{equation}

The energy measured by the static local observer is instead
\begin{equation}
E_{\rm loc}
=
\sqrt{
m^{2}c^{4}
+
p_{\rm loc}^{2}c^{2}
},
\label{eq:energy_local_exact}
\end{equation}
and therefore
\begin{equation}
E_{\infty}
=
N(r)E_{\rm loc}.
\label{eq:redshift_energy_relation}
\end{equation}
Eq. \eqref{eq:redshift_energy_relation} is the standard gravitational redshift relation between the locally measured energy and the conserved energy referred to an observer at spatial infinity.

In the nonrelativistic regime, $p_{\rm loc}\ll mc$, the local energy can be approximated as
\begin{equation}
E_{\rm loc}
\simeq
mc^{2}
+
\frac{p_{\rm loc}^{2}}{2m}.
\label{eq:local_energy_nonrelativistic}
\end{equation}
Consequently, the conserved nonrelativistic energy becomes
\begin{equation}
E_{\infty}^{\rm NR}
\simeq
N(r)
\left(
mc^{2}
+
\frac{p_{\rm loc}^{2}}{2m}
\right).
\label{eq:energy_schwarzschild_nonrelativistic}
\end{equation}
This expression will be used in the following section to construct the Maxwell--Boltzmann number density and the corresponding Saha equation in the Schwarzschild spacetime.

\section{Results and discussion}
\label{sec:results}

\subsection{Saha equation in the Schwarzschild spacetime}
\label{sec:saha_schwarzschild}

We now consider a nonrelativistic particle species \(i\), with degeneracy factor \(g_i\), rest mass \(m_i\), and chemical potential \(\mu_{i,\infty}\). In the Maxwell--Boltzmann approximation, its local number density can be written in terms of asymptotic thermodynamic variables as
\begin{equation}
n_i(r)
=
\frac{g_i}{(2\pi\hbar)^{3}}
\int d^{3}p_{\rm loc}
\exp
\left[
-\frac{
E_{i,\infty}^{\rm NR}
-
\mu_{i,\infty}
}{
k_{\rm B}T_{\infty}
}
\right].
\label{eq:number_density_general}
\end{equation}
Substituting Eq.~\eqref{eq:energy_schwarzschild_nonrelativistic}, we obtain
\begin{equation}
n_i(r)
=
g_i
\left[
\frac{
m_i k_{\rm B}T_{\infty}
}{
2\pi\hbar^{2}N(r)
}
\right]^{3/2}
\exp
\left[
\frac{
\mu_{i,\infty}
-
m_i c^{2}N(r)
}{
k_{\rm B}T_{\infty}
}
\right].
\label{eq:number_density_schwarzschild}
\end{equation}
Using Eq.~\eqref{eq:number_density_schwarzschild} for each particle species, the Saha equation in the Schwarzschild spacetime reads as
\begin{equation}
\frac{
n_p(r)n_e(r)
}{
n_{H_n}(r)
}
=
G_{(e,p,n)}
\left[
\frac{
m_{\rm red}k_{\rm B}T_{\infty}
}{
2\pi\hbar^{2}N(r)
}
\right]^{3/2}
\exp
\left[
-\frac{
N(r)\varepsilon_n
}{
k_{\rm B}T_{\infty}
}
\right].
\label{eq:saha_schwarzschild}
\end{equation}

In global thermal equilibrium, the local temperature satisfies the Tolman--Ehrenfest relation \cite{tolman1930temperature,tolman1935thermal},
\begin{equation}
T_{\rm loc}(r)N(r)
=
T_{\infty},
\label{eq:tolman_relation}
\end{equation}
while the local chemical potential satisfies the Klein relation \cite{klein1949thermodynamical},
\begin{equation}
\mu_{i,\rm loc}(r)N(r)
=
\mu_{i,\infty}.
\label{eq:klein_relation}
\end{equation}
Therefore, by imposing the Tolman--Ehrenfest equilibrium condition, Eq.~\eqref{eq:saha_schwarzschild} can be rewritten entirely in terms of locally measured quantities as
\begin{equation}
\frac{
n_p(r)n_e(r)
}{
n_{H_n}(r)
}
=
G_{(e,p,n)}
\left[
\frac{
m_{\rm red}k_{\rm B}T_{\rm loc}(r)
}{
2\pi\hbar^{2}
}
\right]^{3/2}
\exp
\left[
-\frac{
\varepsilon_n
}{
k_{\rm B}T_{\rm loc}(r)
}
\right].
\label{eq:saha_schwarzschild_local}
\end{equation}
Hence, the local Saha equation preserves its standard flat-spacetime form. The Schwarzschild gravitational field affects the ionization equilibrium through the radial dependence of the Tolman--Ehrenfest temperature and through the redshift relating local quantities to those measured by an asymptotic observer. This result is consistent with the equivalence principle, since the intrinsic ionization energy remains unchanged in a sufficiently small local inertial frame.

In what follows, we discuss the main consequences of the Schwarzschild geometry for the ionization equilibrium. Since the gravitational effects are entirely encoded in the lapse function, Eq. \eqref{eq:schwarzschild_lapse}, it is convenient to introduce the dimensionless radial coordinate
\begin{equation}
x\equiv\frac{r}{r_{\rm s}},
\qquad
r_{\rm s}\equiv\frac{2GM}{c^{2}},
\label{eq:dimensionless_radius}
\end{equation}
for which
\begin{equation}
N(x)
=
\sqrt{1-\frac{1}{x}},
\qquad x>1.
\label{eq:lapse_dimensionless}
\end{equation}
This parametrization makes the results independent of the mass of the central object when distances are expressed in units of the Schwarzschild radius.

\subsection{Gravitational redshift of the ionization energy}

\begin{figure}[t]
    \centering
    \includegraphics[width=0.48\textwidth]{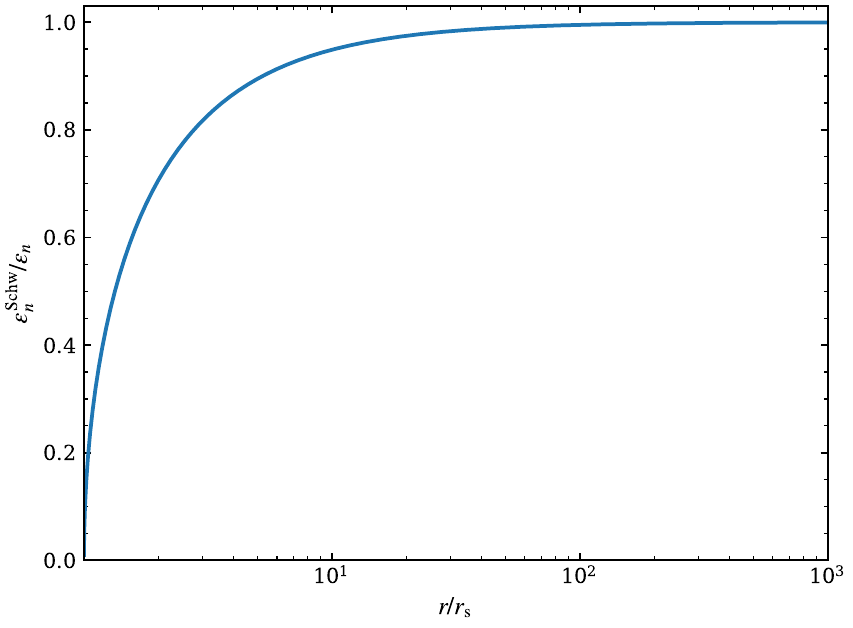}
    \caption{Normalized ionization energy measured by an asymptotic observer as a function of the dimensionless Schwarzschild radial coordinate \(r/r_{\rm s}\).}
    \label{fig:binding_energy}
\end{figure}

From the point of view of an observer located at spatial infinity, the ionization energy of the \(n\)th atomic level is redshifted according to
\begin{equation}
\varepsilon_n^{\rm Schw}(r)
=
N(r)\varepsilon_n.
\label{eq:redshifted_binding_results}
\end{equation}
Figure~\ref{fig:binding_energy} shows the normalized ionization energy as a function of \(r/r_{\rm s}\). As expected, the asymptotically measured energy approaches the standard energy for \(r/r_{\rm s}\gg1\), whereas it decreases monotonically as the Schwarzschild radius is approached,
\begin{equation}
\frac{\varepsilon_n^{\rm Schw}}{\varepsilon_n}
\longrightarrow 0
\qquad
\text{for}
\qquad
r\rightarrow r_{\rm s}^{+}.
\end{equation}

This behavior must not be interpreted as a local suppression of the ionization energy. Rather, it represents the gravitational redshift of the transition energy with respect to an observer at infinity. The intrinsic ionization energy measured in a sufficiently small local inertial frame remains equal to \(\varepsilon_n\), in accordance with the equivalence principle.

\subsection{Gravitational modification of the Saha ratio}

A useful way of quantifying the gravitational contribution is to compare the Schwarzschild Saha ratio with its flat-spacetime counterpart at the same asymptotic temperature \(T_\infty\). For this purpose, we define
\begin{equation}
R_{\rm Schw}(r,T_\infty)
\equiv
\frac{n_p(r)n_e(r)}{n_{H_n}(r)},
\end{equation}
where $R_{\rm Schw}$ is given by Eq. \eqref{eq:saha_schwarzschild}. The corresponding flat-spacetime expression ($R_{\rm flat}$), evaluated at the same \(T_\infty\), is given by Eq. \eqref{eq:saha_standard}. The relative modification is therefore
\begin{equation}
\mathcal{R}(r,T_\infty)
\equiv
\frac{R_{\rm Schw}}{R_{\rm flat}}
=
N(r)^{-3/2}
\exp\left[
\frac{\varepsilon_n}{k_{\rm B}T_\infty}
\left(1-N(r)\right)
\right].
\label{eq:saha_ratio_relative}
\end{equation}

\begin{figure}[t]
    \centering
    \includegraphics[width=0.48\textwidth]{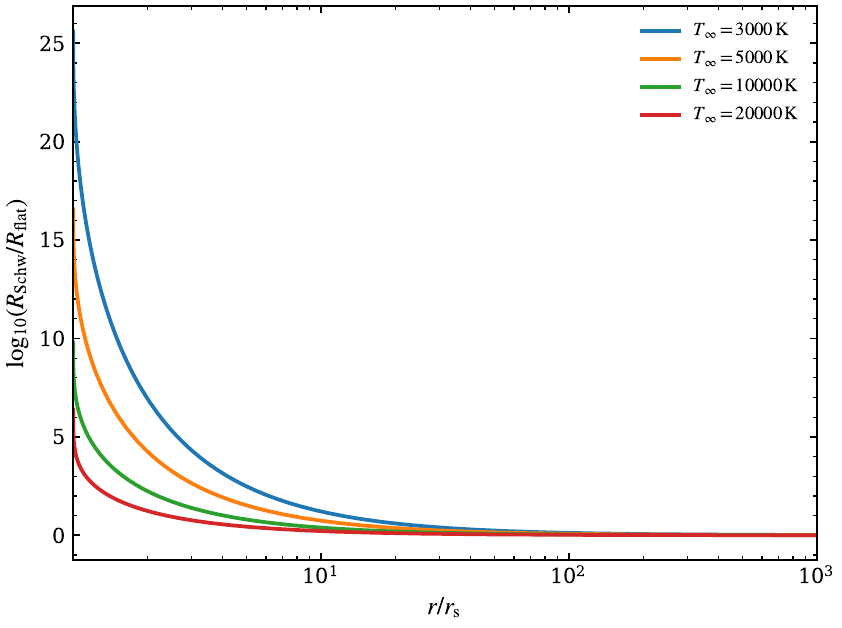}
    \caption{Relative modification of the Saha ionization ratio, expressed as \(\log_{10}(R_{\rm Schw}/R_{\rm flat})\), as a function of \(r/r_{\rm s}\) for different values of the asymptotic temperature \(T_\infty\).}
    \label{fig:saha_ratio}
\end{figure}

Figure~\ref{fig:saha_ratio} displays $\log_{10}\mathcal{R}$ for different values of \(T_\infty\). The gravitational enhancement of the ionization ratio becomes increasingly pronounced as the Schwarzschild radius is approached. Two contributions are responsible for this behavior. The first arises from the translational factor \(N^{-3/2}\), whereas the second originates from the exponential dependence on the redshifted ionization energy. The latter contribution is particularly important because the exponential argument becomes larger at lower temperatures. Hence, the relative gravitational effect is stronger for colder plasmas when the comparison is performed at fixed \(T_\infty\). In the asymptotically flat limit, $N(r)\rightarrow1$, 
and Eq.~\eqref{eq:saha_ratio_relative} correctly gives $\mathcal{R}\rightarrow1$.

\subsection{Ionization fraction}

The physical consequences of the previous result can be expressed more directly in terms of the ionization fraction. For a pure hydrogen plasma satisfying charge neutrality, we write $n_e=n_p=x_e n_{\rm H}$ and $n=(1-x_e)n_{\rm H}$, where \(n_{\rm H}\) denotes the total hydrogen number density. The Saha equation then becomes
\begin{equation}
\frac{x_e^2}{1-x_e}
=
\frac{R_{\rm Schw}(r,T_\infty)}
{n_H}.
\label{eq:xe_saha_results}
\end{equation}

\begin{figure}[ht]
    \centering
    \includegraphics[width=0.48\textwidth]{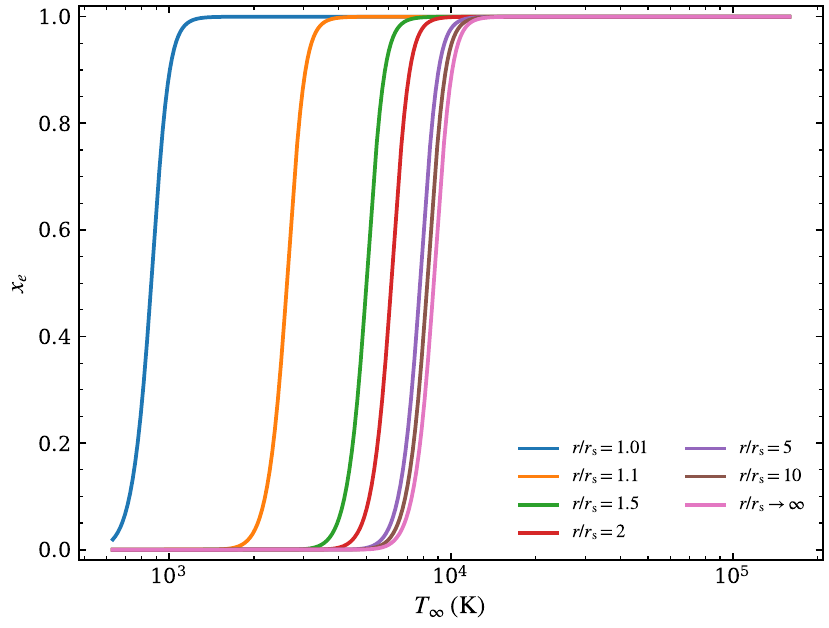}
    \caption{Ionization fraction \(x_e\) as a function of the asymptotic temperature \(T_\infty\) for different values of \(r/r_{\rm s}\). The hydrogen number density is fixed at $n_{\rm H}=10^{20}\,{\rm m}^{-3}$.}
    \label{fig:ionization_fraction}
\end{figure}

Figure~\ref{fig:ionization_fraction} shows the evolution of \(x_e\) with \(T_\infty\) for several radial positions. Far from the compact object, the flat-spacetime Saha curve is recovered. As the system approaches the Schwarzschild radius, the transition between the predominantly neutral and predominantly ionized regimes is displaced toward lower values of the asymptotic temperature. It is worth stressing that this shift can also be explained from the Tolman--Ehrenfest law, Eq. \eqref{eq:tolman_relation}. At a fixed \(T_\infty\), the local temperature increases as \(N(r)\) decreases. Thus, a plasma located deeper in the gravitational potential corresponds locally to a thermodynamically hotter system, which favors ionization. The displacement observed in Fig.~\ref{fig:ionization_fraction} therefore reflects the gravitational relation between local and asymptotic temperatures rather than a modification of the local atomic physics.

\subsection{Relativistic classical counterpart}
\label{sec:relativistic_counterpart}

For completeness, the nonrelativistic construction above can be embedded in a relativistic classical ideal gas treatment. We follow the same procedure as before, but now retain the exact Killing energy in Eq. \eqref{eq:energy_infinity_exact}. For a classical, nondegenerate species $i$, the number density is therefore
\begin{equation}
\begin{split}
n_i(r)
={}&
\frac{g_i}{(2\pi\hbar)^3}
\int d^3p_{\rm loc}\,
\exp\!\left[
-\frac{N(r)\sqrt{m_i^2c^4+p_{\rm loc}^2c^2}}
{k_{\rm B}T_\infty}
\right]
\\
&\times
\exp\!\left(\frac{\mu_{i,\infty}}{k_{\rm B}T_\infty}\right).
\end{split}
\label{eq:relativistic_density_integral}
\end{equation}
The Maxwell--J\"uttner momentum integral can be evaluated in terms of the modified Bessel function of the second kind, $K_2$ \cite{deGroot1980,das2018saha}. The result is
\begin{equation}
n_i(r)
=
\frac{g_i m_i^2 c k_{\rm B}T_\infty}
{2\pi^2\hbar^3N(r)}
K_2\!\left(\frac{m_i c^2N(r)}{k_{\rm B}T_\infty}\right)
\exp\!\left(\frac{\mu_{i,\infty}}{k_{\rm B}T_\infty}\right).
\label{eq:relativistic_density_schwarzschild}
\end{equation}
Introducing
\begin{equation}
z_i(r)=\frac{m_i c^2N(r)}{k_{\rm B}T_\infty},
\label{eq:relativistic_bessel_argument}
\end{equation}
and imposing chemical equilibrium, Eq.~\eqref{eq:chemical_equilibrium}, the fugacity factors cancel and the relativistic classical Saha equation written in terms of the lapse and asymptotic variables becomes
\begin{equation}
\frac{n_p(r)n_e(r)}{n_{H_n}(r)}
=
G_{(e,p,n)}
\frac{c k_{\rm B}T_\infty}{2\pi^2\hbar^3N(r)}
\frac{m_p^2m_e^2}{m_{H_n}^2}
\frac{K_2(z_p)K_2(z_e)}{K_2(z_{H_n})}.
\label{eq:saha_relativistic_schwarzschild}
\end{equation}

In this stage, we can impose the Tolman--Ehrenfest relation, Eq.~\eqref{eq:tolman_relation}, so that the arguments become $z_i=m_i c^2/[k_{\rm B}T_{\rm loc}(r)]$, and Eq.~\eqref{eq:saha_relativistic_schwarzschild} assumes the local form
\begin{equation}
\frac{n_p n_e}{n_{H_n}}
=
G_{(e,p,n)}
\frac{c k_{\rm B}T_{\rm loc}}{2\pi^2\hbar^3}
\frac{m_p^2m_e^2}{m_{H_n}^2}
\frac{K_2(z_p)K_2(z_e)}{K_2(z_{H_n})}~.
\label{eq:saha_relativistic_local}
\end{equation}
Thus, the explicit Schwarzschild dependence again disappears in local variables. Equation~\eqref{eq:saha_relativistic_local} should be understood as a relativistic classical ideal gas counterpart of the Saha equation, rather than as a complete description of a relativistic plasma.

The size of the kinematic correction follows directly from the standard large-$z$ expansion of the modified Bessel function \cite{deGroot1980,das2018saha},
\begin{equation}
K_2(z)
=
\sqrt{\frac{\pi}{2z}}e^{-z}
\left(1+\frac{15}{8z}+\mathcal{O}(z^{-2})\right).
\label{eq:k2_large_z}
\end{equation}
Relative to the leading nonrelativistic result, this gives
\begin{equation}
\frac{\Delta R}{R_{\rm NR}}
\simeq
\frac{15k_{\rm B}T_{\rm loc}}{8c^2}
\left(\frac{1}{m_p}+\frac{1}{m_e}-\frac{1}{m_{H_n}}\right)
\simeq
\frac{15}{8}\frac{k_{\rm B}T_{\rm loc}}{m_ec^2},
\label{eq:relativistic_saha_correction}
\end{equation}
Here $\Delta R\equiv R_{\rm rel}-R_{\rm NR}$, where $R_{\rm rel}$ denotes the relativistic ionization ratio in Eq.~\eqref{eq:saha_relativistic_local} and $R_{\rm NR}$ is its nonrelativistic limit. Thus, Eq.~\eqref{eq:relativistic_saha_correction} gives the fractional kinematic correction to the Saha ratio. The last expression uses $m_p,m_{H_n}\gg m_e$. For example, the correction is approximately $3.7\times10^{-6}$ at $k_{\rm B}T_{\rm loc}=1\,{\rm eV}$ and $3.7\times10^{-5}$ at $10\,{\rm eV}$. It is therefore negligible throughout the temperature range in which an appreciable neutral hydrogen abundance is expected.

At temperatures high enough for this correction to become substantial, $k_{\rm B}T_{\rm loc}\sim m_ec^2$, hydrogen is already predominantly ionized. In that regime, electron--positron pair creation, Fermi--Dirac statistics, plasma screening, and the survival of atomic bound states must also be considered. Moreover, obtaining such local temperatures from the Tolman--Ehrenfest relation at fixed $T_\infty$ requires approaching the horizon extremely closely. Since the proper acceleration of a static observer diverges as $r\rightarrow r_{\rm s}$, a static test atmosphere cannot be regarded as realistic arbitrarily near the horizon. These restrictions delimit the physical interpretation of Eq.~\eqref{eq:saha_relativistic_local}, while its recovery of the nonrelativistic result provides a useful consistency check of the formulation.

\section{Conclusion} 
\label{sec:conclusion}

We have developed a formulation of the Saha ionization equilibrium for a dilute hydrogen plasma in the exterior Schwarzschild spacetime. The analysis was based on the conserved Killing energy of a massive particle, from which the corresponding nonrelativistic expression was obtained and subsequently introduced into the Maxwell--Boltzmann distribution. The resulting ionization relation contains an explicit dependence on the Schwarzschild lapse function when expressed in terms of quantities referred to an observer at spatial infinity. In this description, both the characteristic ionization energy and the equilibrium ionization ratio are gravitationally redshifted. The numerical results show that these effects become increasingly significant as the plasma is located deeper in the gravitational potential, while the standard flat-spacetime behavior is recovered in the asymptotic region.

A central result of the present analysis is that this gravitational dependence does not represent an intrinsic modification of the local ionization energy. Once the Tolman--Ehrenfest relation for the temperature and the Klein condition for the chemical potential are consistently imposed, the ionization equation written in terms of locally measured thermodynamic variables recovers the standard Saha form. The displacement of the ionization transition toward lower asymptotic temperatures near the Schwarzschild radius therefore reflects the gravitational relation between local and distant observers rather than a change in the underlying atomic physics.

For completeness, we have also embedded the nonrelativistic calculation in a relativistic classical ideal gas formulation based on the Maxwell--J\"uttner distribution. Its large-mass expansion recovers the nonrelativistic Saha equation and gives a leading fractional correction of order $k_{\rm B}T_{\rm loc}/(m_ec^2)$. This correction is negligible when neutral hydrogen is appreciably abundant, while the regime in which it becomes significant requires additional effects such as pair production and quantum statistics.

This distinction is particularly important in interpreting ionization processes in strong gravitational fields. The Schwarzschild geometry modifies the connection between locally measured and asymptotically observed quantities, while the equilibrium microphysics remains locally equivalent to that of flat spacetime, as required by the equivalence principle. The present framework therefore provides a simple thermodynamic description of ionization equilibrium in static, spherically symmetric gravitational fields and may serve as a starting point for extensions involving more realistic plasma environments, nonequilibrium effects, or additional spacetime geometries. Natural extensions include the investigation of thermal and
boundary-induced quantum effects, such as the thermal Casimir effect in
compact-star spacetimes \cite{de2025stefan}; the incorporation of
non-extensive statistics, along the lines of
Refs. \cite{sales2022non,sales2023constraint,sales2025possible}; the
generalization to rotating (Kerr) spacetimes, relevant for realistic
astrophysical compact objects; and the inclusion of nonequilibrium kinetic
corrections analogous to those required for an accurate description of
cosmological recombination
\cite{peebles1968recombination,seager1999new,chluba2006free,
chluba2011towards,dubrovich2005recombination}.

\begin{acknowledgments}
LLS and ARQ thank the Para\'iba State Research Foundation (FAPESQ) for financial support.  ARQ also acknowledges the support of CNPq under process number 306884/2026-7. 
\end{acknowledgments}

\bibliography{mybibfile}

@article{Saha1920,
	title={Ionisation in the solar chromosphere},
	author={Saha, MN},
	journal={Nature},
	volume={105},
	number={2634},
	pages={232--233},
	year={1920},
	publisher={Nature Publishing Group UK London}
}

@book{deGroot1980,
  title={Relativistic Kinetic Theory: Principles and Applications},
  author={de Groot, Sybren R. and van Leeuwen, Willem A. and van Weert, Chushiro G.},
  year={1980},
  publisher={North-Holland},
  address={Amsterdam}
}

@article{das2018saha,
  title={Saha equation for the photo-ionization of hydrogen atoms in partially ionized relativistic hydrogen plasma and the effect of gravity on the binding of hydrogen atoms in Rindler space},
  author={Das, Sanchita and Chakrabarty, Somenath},
  journal={Open Access J. Phys.},
  volume={2},
  number={3},
  pages={25--30},
  year={2018},
  doi={10.22259/2637-5826.0203004},
  eprint={1801.06774},
  archivePrefix={arXiv},
  primaryClass={gr-qc}
}

@article{gangopadhyay2021hundred,
	title={Hundred years of the Saha equation and astrophysics},
	author={Gangopadhyay, Gautam},
	journal={EPJ - Special Topics},
	volume={230},
	pages={495--503},
	year={2021},
	publisher={Springer}
}

@book{peebles2020principles,
	title={Principles of physical cosmology},
	author={Peebles, Phillip James Edwin},
	year={2020},
	publisher={Princeton university press}
}

@book{dodelson2024modern,
	title={Modern cosmology},
	author={Dodelson, Scott and Schmidt, Fabian},
	year={2024},
	publisher={Elsevier}
}

@article{peebles1968recombination,
	title={Recombination of the primeval plasma},
	author={Peebles, PJE},
	journal={ApJ},
	volume={153},
	pages={1},
	year={1968}
}

@article{sales2022non,
	title={Non-Gaussian effects of the Saha’s ionization in the early universe},
	author={Sales, LL and Carvalho, FC and Bento, EP and Souza, HTCM},
	journal={Eur. Phys. J. C},
	volume={82},
	number={1},
	pages={54},
	year={2022},
	publisher={Springer}
}

@article{sales2023constraint,
  title={Constraint on the chemical potentials of hydrogen and proton in recombination},
  author={Sales, LL and Carvalho, FC and Souza, HTCM},
  journal={Eur. Phys. J. C},
  volume={83},
  number={6},
  pages={466},
  year={2023},
  publisher={Springer}
}

@article{sales2025possible,
  title={A possible correction of the Saha curve for non-equilibrium states},
  author={Sales, LL and Carvalho, FC and Souza, H TCM},
  journal={Phys. Lett. B},
  pages={140011},
  year={2025},
  publisher={Elsevier}
}

@article{sales2024non,
  title={Non-gaussian Saha’s ionization in Rindler spacetime and the equivalence principle},
  author={Sales, LL and Carvalho, FC},
  journal={Eur. Phys. J. C},
  volume={84},
  number={7},
  pages={671},
  year={2024},
  publisher={Springer}
}

@article{seager1999new,
	title={A new calculation of the recombination epoch},
	author={Seager, Sara and Sasselov, Dimitar D and Scott, Douglas},
	journal={ApJ},
	volume={523},
	number={1},
	pages={L1},
	year={1999},
	publisher={IOP Publishing}
}

@article{dubrovich2005recombination,
	title={Recombination dynamics of primordial hydrogen and helium (He I) in the universe},
	author={Dubrovich, VK and Grachev, SI},
	journal={Astron. Lett.},
	volume={31},
	pages={359--364},
	year={2005},
	publisher={Springer}
}

@article{chluba2011towards,
	title={Towards a complete treatment of the cosmological recombination problem},
	author={Chluba, J and Thomas, RM},
	journal={MNRAS},
	volume={412},
	number={2},
	pages={748--764},
	year={2011},
	publisher={Blackwell Publishing Ltd Oxford, UK}
}

@article{chluba2006free,
	title={Free-bound emission from cosmological hydrogen recombination},
	author={Chluba, Jens and Sunyaev, RA},
	journal={A\&A},
	volume={458},
	number={2},
	pages={L29--L32},
	year={2006},
	publisher={EDP Sciences}
}

@article{tolman1930temperature,
  title={Temperature equilibrium in a static gravitational field},
  author={Tolman, Richard C and Ehrenfest, Paul},
  journal={Phys. Rev.},
  volume={36},
  number={12},
  pages={1791},
  year={1930},
  publisher={APS}
}

@article{tolman1935thermal,
  title={Thermal equilibrium in a general gravitational field},
  author={Tolman, Richard C},
  journal={Proc. Natl. Acad. Sci. U. S. A.},
  volume={21},
  number={6},
  pages={321--326},
  year={1935}
}

@article{lee2020hyrec,
  title={HYREC-2: a highly accurate sub-millisecond recombination code},
  author={Lee, Nanoom and Ali-Ha{\"\i}moud, Yacine},
  journal={Physical Review D},
  volume={102},
  number={8},
  pages={083517},
  year={2020},
  publisher={APS}
}

@article{de2025stefan,
  title={Stefan-Boltzmann Law and Thermal Casimir Effect in Neutron Star Spacetime via Thermo Field Dynamics},
  author={de Farias, Klecio EL and Anacleto, Marcos A and Batista, Rafael A and Brevik, Iver and Brito, Francisco A and Passos, Eduardo and Queiroz, Amilcar R and Sales, L{\'a}zaro L},
  journal={arXiv preprint arXiv:2512.15610},
  year={2025}
}

@article{Santiago_2019,
doi = {10.1088/1361-6404/aaff1c},
url = {https://doi.org/10.1088/1361-6404/aaff1c},
year = {2019},
month = {feb},
publisher = {IOP Publishing},
volume = {40},
number = {2},
pages = {025604},
author = {Santiago, Jessica and Visser, Matt},
title = {Tolman temperature gradients in a gravitational field},
journal = {Eur. J. Phys.}
}

@article{klein1949thermodynamical,
  title={On the thermodynamical equilibrium of fluids in gravitational fields},
  author={Klein, O},
  journal={Rev. Mod. Phys.},
  volume={21},
  number={3},
  pages={531},
  year={1949},
  publisher={APS}
}

@article{lima2019thermodynamic,
  title={Thermodynamic equilibrium in general relativity},
  author={Lima, Jose Ademir Sales de and Del Popolo, A and Plastino, AR},
  journal={Phys. Rev. D},
  volume={100},
  number={10},
  pages={104042},
  year={2019},
  publisher={APS}
}

\end{document}